# Classical analog of Stückelberg interferometry with two coupled mechanical resonators

Hao Fu,[1] Zhi-cheng Gong,[1] Tian-hua Mao,[1] Chang-pu Sun,[2,3] Su Yi,[4] Yong Li,[2,3,*] & Geng-yu Cao[1,*]

[1] State Key Laboratory of Magnetic Resonance and Atomic and Molecular Physics, Wuhan Institute of Physics and Mathematics, Chinese Academy of Sciences, Wuhan 430071, P. R. China

[2] Beijing Computational Science Research Center, Beijing 100193, China

[3] Synergetic Innovation Center of Quantum Information and Quantum Physics, University of Science and Technology of China, Hefei, Anhui 230026, China

[4] Institute of Theoretical Physics, Chinese Academy of Sciences, Beijing 100190, China

*Correspondence and requests for materials should be addressed to Y. L. (liyong@csrc.ac.cn) or G.-Y. C. (gycao@wipm.ac.cn)

**Abstract**

Coupled nanomechanical resonators have recently attracted much attention for both fundamental studies in physics and broad applications in high-precession detection or sensing. By studying the Landau-Zener transitions and Rabi oscillation of two coupled resonators, it has been shown that such a two-mode system acts as a classical two-level system bearing the analog to a quantum mechanical two-level one. Here we construct a Stückelberg interferometer with two coupled cantilevers by driving the system through the avoided crossing twice. By measuring the non-adiabatic phase acquired at the Landau-Zener transition, we unveil an in-depth analog between the two-mode and quantum two-level systems. Our study opens up new opportunities for constructing interferometers with classical devices.

Following the rapid development of nanotechnologies, coupled nanomechanical resonators have recently attracted much attention for both fundamental studies in physics [1, 2, 3] and broad applications in high-precession detection or sensing [4, 5, 6, 7, 8, 9]. Mechanical couplings between different resonators [10, 11, 12] and between different oscillation modes of the same resonator [13, 14, 15] have been demonstrated. In particular, coherent manipulation of the mechanical oscillations through the Rabi-like oscillation is realized for two strongly coupled mechanical resonators [15, 16, 17]. It is also shown that a classical analog of the Landau-Zener (LZ) transitions in quantum two-level systems can be realized by driving the two mechanical modes of a single nanomechanical beam through the avoid crossing [18]. To further investigate the coherence of the coupled resonators across the LZ transition, one may drive the system through the avoided crossing twice. The first passage coherently splits mechanical oscillations between two resonators and the second one creates their superpositions in both resonators through recombination. As a result, a double-passage process leads to the well-known Stückelberg oscillation, an interference fringe in time domain [19, 20]. Quantum mechanically, the Stückelberg oscillations have been demonstrated in a wide variety of systems [21, 22, 23, 24]. In particular, periodic passages through the avoided crossing lead to the Landau-Zener-Stückelberg-Majorana interference which can be used to determine the energy level diagram and to measure coherence time of superconducting qubits [25, 26, 27, 28, 29].

In this work, we report the realization of a Stückelberg interferometer with two coupled single-crystal-silicon cantilevers. Focusing on the fundamental flexural mode of each cantilever, the frequency of one of these modes can be tuned by placing the corresponding cantilever inside a fiber-based Fabry-Pérot cavity and time-resolvedly controlling the pumping power of the cavity. To facilitate the construction of the Stückelberg interferometer, we first explore the coupling between these two modes by driving the system through the avoided crossing which induces a Landau-Zener-like transition. We then demonstrate that the two-mode system can be coherently manipulated by realizing the Rabi-like oscillation. Finally, we realize the Stückelberg interferometer by driving the system over the avoided crossing twice and obtain the interference fringes of the mechanical oscillations in one cantilever. In addition, we determine the Stokes phase acquired during the non-adiabatic transition when passing through the avoid crossings.

**Results**

**Tunable two-mode optomechanical system.** As illustrated in Fig. 1a, the system under investigation consists of two single-crystal-silicon cantilevers connected to the same thin overhang extended $20 \pm 5$ μm out the insulator substrate. Here, we focus on the fundamental flexural modes originating from the displacement $x_i$ ($i = 1,2$) of the cantilevers. Due to the inhomogeneous etching of the substrate, the intrinsic frequencies of cantilevers are different even though their geometrics and effective mass are essentially the same. In order to tune the intrinsic frequency of the cantilever 1, we insert it into a fiber-based Fabry-Pérot cavity (Fig. 1a); the cavity is then pumped from one side by a 1064 nm Nd-YAG laser with the laser power $P$. In this configuration, the fibers and cantilever 1 form a membrane-in-the-middle optomechanical system [30, 31, 32], in which the dispersive coupling between the cantilever and the cavity field gives rise to a harmonic trapping to the cantilever corresponding to an effective frequency $\sqrt{gP}$, where $g$ is the optical trapping strength. Consequently, the effective frequency of the cantilever 1, $\omega_1(P) = \sqrt{\omega_1^2(0) + gP}$, becomes laser-power dependent [33], where $\omega_1(0)$ is the intrinsic frequency of the cantilever 1 in the absence of pumping laser; while the intrinsic frequency of the cantilever 2, $\omega_2$, remains unaffected. Furthermore, the coupling between the cantilevers through the motion-induced stress inside the overhang results in the hybridization of the two fundamental flexural modes and gives rise to two normal modes $X_\alpha$ ($\alpha = +, -$) which are linear combinations of $x_1$ and $x_2$ [12]. Correspondingly, the frequencies of these normal modes are denoted as $\omega_+(P)$ for the out-of-phase mode and $\omega_-(P)$ the in-phase mode (Fig. 1b).

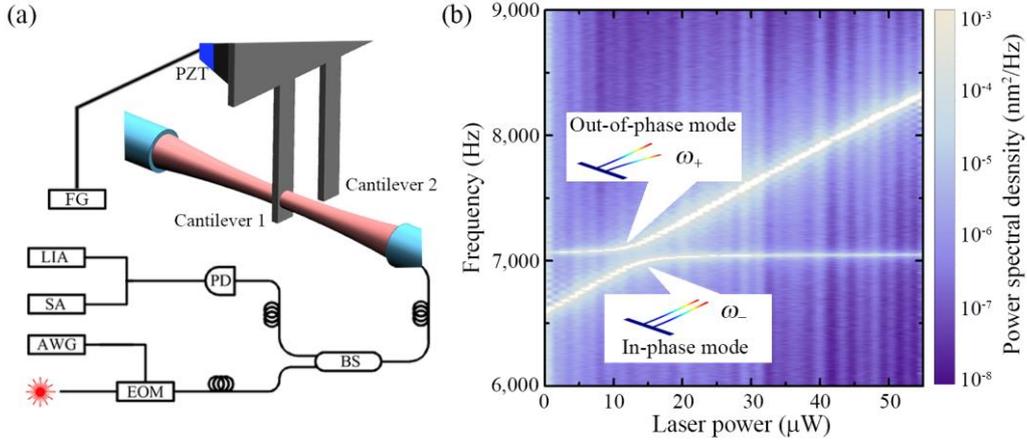

**Figure 1 | Two-mode optomechanical system.** (a) Schematic drawing of coupled-cantilever-based two-mode optomechanical system. The coupled mechanical resonators consist of two 220-μm-long, 10-μm-wide, and 220-nm-thick single-crystal-silicon cantilevers with a separation of 15 μm. Both cantilevers structurally connect to the same thin overhang extended 20 μm over a thick $SiO_2$ substrate. A PZT driven by a function generator (FG) is attached to the substrate of the cantilevers to excite the mechanical motions of the system. The 4.5-mm-long fiber cavity is formed by two gold-coated fiber ends. Light inside the cavity is focused by a micro-lens to the waist radius of 3 μm in the middle. (b) Oscillation power spectral density of the coupled cantilevers versus laser power. The two resonances correspond to the oscillations of the in-phase and out-of-phase motions of the coupled cantilevers as shown in the insets. A normal-mode splitting is observed with the avoided crossing at the laser power of $P_{AC} = 12.9\ \mu W$.

When the laser power is swept, the avoided crossing of the two normal modes is mapped out in the thermal oscillation spectrum of cantilevers (see Fig. 1b and Methods), in which a minimal frequency splitting of $\Delta/2\pi = 156.8$ Hz is observed at the laser power $P_{AC} = 12.9$ μW. It can be further determined from the normal-mode spectrum that the intrinsic frequencies of the cantilevers and the optical trapping strength are, respectively, $\omega_1(0)/2\pi = 6,600$ Hz, $\omega_2/2\pi = 7,069$ Hz and $g/(2\pi)^2 = 4.72 \times 10^5 \text{Hz}^2/\mu\text{W}$. Also, in order to obtain Fig. 1b, we have tuned the length of the fiber-based cavity such that the mechanical damping rates of the normal modes are independent of the laser power [12, 33]. In fact, it is measured that both the normal modes have essentially the same damping rate $\gamma/2\pi = 3.85$ Hz, which clearly indicates that the system is in the strong coupling regime as $\gamma \ll \Delta$ is satisfied. However, far away from the avoided crossing, the oscillation of the cantilever 1 is dominated by the in-phase mode at low laser power and by the out-of-phase mode at high laser power.

**Landau-Zener-like transitions of mechanical oscillations.** The two-mode optomechanical system under consideration is analogous to a quantum spin-1/2 system subjected to a longitudinal magnetic field $\varepsilon(P) \equiv \omega_1(P) - \omega_2$ and a transverse magnetic field $\Delta$. Therefore, driving a classical two-mode system through the avoided crossing may also induce a Landau-Zener-like transition of mechanical oscillations between different normal modes [18]. The experimental sequence for demonstrating such a transition is depicted in Fig. 2a (see Methods). Specifically, the system is prepared by resonantly exciting the in-phase mode at the laser power $P_0 = 0.66$ μW such that about 98% of the mechanical energy is initialized to the cantilever 1. Then from $t = 0$, we linearly ramp up the laser power to $P_f = 54.0$ μW within time $t_f$. This process may lead part of the mechanical energy transferring from the in-phase mode to the out-of-phase mode. Here the LZ transition probability is the fraction of the mechanical energy transferred from the in-phase mode to the out-of-phase mode, i.e., $p_{\text{LZ}} = A_+(t_f)^2/A_-(0)^2$, where $A_\alpha$ is the amplitude of the $\alpha$-th normal mode. Figure 2b shows the transition probability $p_{\text{LZ}}$ as a function of the sweeping rate $v \equiv \frac{d\varepsilon(P)}{dt}\big|_{P=P_{AC}}$ which is defined at the avoided crossing. As can be seen, the measured transition probability ranges from zero to unit, indicating that, by tuning the sweeping rate, the coupled cantilevers allow adiabatic transfer, coherent splitting and diabatic transfer for the oscillation state. In particular, a 50/50-splitting is achieved at the sweeping rate $v/2\pi = 342$ kHz/s. More remarkably, as shown in Fig. 2b, the transition probability also satisfies the standard LZ formula, $p_{LZ} = \exp(-\pi\Delta^2/2|v|)$, which justifies it as a classical analogy of the LZ transition. Finally, we remark that the measured transition probabilities are also in good agreement with our full numerical simulations for the classical two-level system (see Supplementary Information for details).

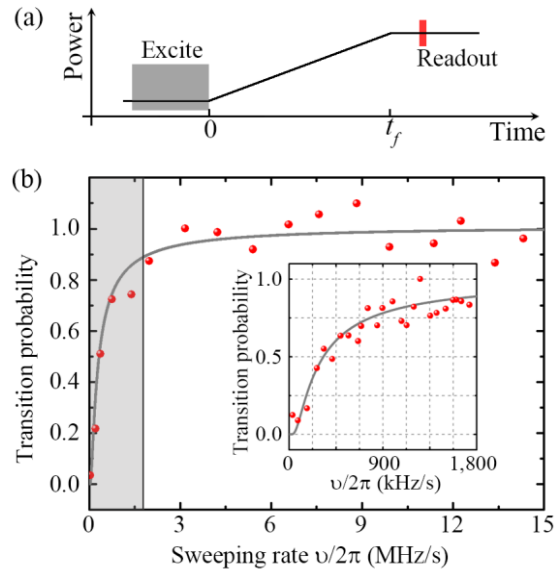

**Figure 2 | LZ transition between mechanical modes.** (a) Optical driving scheme. After a single passage transition, the laser power is hold for optical readout of mechanical oscillations. (b) Transition probability of mechanical oscillations versus sweeping rate. Theoretical result (gray line) calculated from the standard LZ formula is plotted together with the experimental measurement (red dots). The shaded region is showed in detail in the inset.

**Rabi oscillation of the coupled cantilevers.** To facilitate the construction of the Stückelberg interferometer, we shall first show that the two cantilevers can exchange oscillations coherently via the Rabi oscillation. To this aim, as shown in Fig. 3a, we initialize the system by exciting the in-phase mode at $P_0 = 45.5 \, \mu W$ where $\omega_-(P_0) \approx \omega_2$. As a result, almost all of the oscillation energy is localized in the cantilever 2. We then linearly ramped down the laser power to the coupling regime $P_f$ within 15 μs, which concludes the state preparation. Here, since the time used to ramp down the laser power is much smaller than $2\pi/\omega_-(P_0)$, the oscillation energy of the initial state remains being localized in the cantilever 2. We then hold the laser power at $P_f$ and allow the system to evolve for time $\tau$. Finally, the oscillation amplitude of the cantilever 2, $A_2(\tau)$, or equivalently $A_-(\tau)$, is measured 15 ms after we quickly ramp the laser power back to $P_0$.

In Fig. 3b, the Bloch sphere shows the states of the system at selected times. In the context of the quantum spin-1/2 model, the physical process can be understood as follows. The initial state is prepared under a sufficiently large longitudinal magnetic field $\varepsilon(P)$ such that the direction of the total magnetic field, $X_-$, is roughly along the $x_2$ axis. By ramping down the laser power to $P_f$ in the coupling regime, we lower the longitudinal magnetic field at time $t = 0$, which effectively deviates $X_-$ from the $x_2$ axis with angle $\theta$ determined by $\tan\theta = \Delta/\varepsilon(P)$ (see Supplementary Information). After the initialization, the state of the system starts to precess around the $X_-$ axis with the Rabi frequency $\Omega_R = \sqrt{\Delta^2 + \varepsilon^2} = \omega_+ - \omega_-$ such that it becomes a superposition of the oscillations of the two cantilevers. Here, the superposition amplitude for the $i$-th cantilever (or equivalently, the projection of the state along the $x_i$ axis) is directly related to the oscillation energy on that cantilever and thus can be determined by measuring the oscillation amplitude $A_i$ after quickly ramping the laser power back to $P_0$ where two cantilevers are decoupled. Due to the precession of the state (see Fig. 3b), the oscillation amplitudes $A_i$ are clearly periodic functions of $\tau$, representing the coherent oscillation exchange between two cantilevers. We remark that, in terms of the normal modes, the related oscillation amplitudes remain unchanged during the hold time; only a relative phase is acquired.

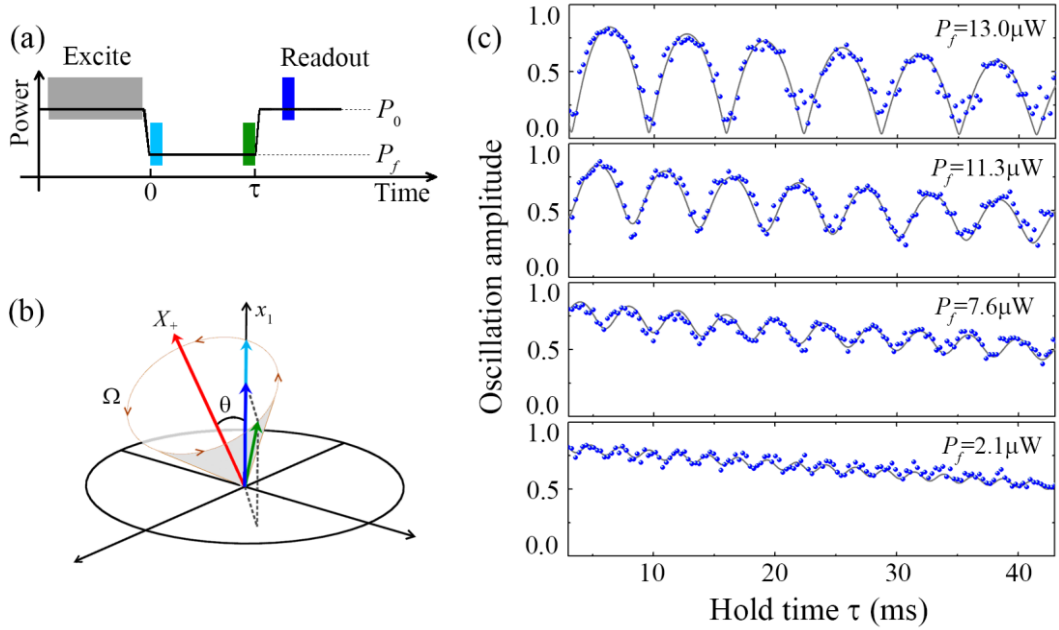

**Figure 3 | Rabi oscillation.** (a) Optical driving scheme. (b) The Bloch vectors showing the states of the system in the Bloch sphere (marked in the same colour in a) at selected times. (c) Rescaled oscillation amplitudes of the cantilever 2, $A_2(\tau)/A_2(0)$, versus $\tau$ for various $P_f$. Measured data are shown as dots and the fit with Eq. (1) is plotted as solid lines.

Figure 3c plots the measurements of the rescaled oscillation amplitude $A_2(\tau)/A_2(0)$ for various $P_f$'s. As can be seen, the maximal value of $A_2(\tau)/A_2(0)$ increases to unit when $P_f$ approaches $P_{AC}$, which reflects the fact that the stronger coupling leads to more intensive exchange. It can also be understood by noting that the precession axis $X_-$ deviates from the $x_2$ axis by $\pi/2$ at $P_f = P_{AC}$.

To gain more insight into the dynamics, we note that the oscillation amplitude of the cantilever 2 is governed by the equation (see Supplementary Information)

$$\frac{A_2(\tau)}{A_2(0)} = e^{-\tau/(2T_d)} \sqrt{1 - \sin^2\theta \sin^2\frac{(\omega_+ - \omega_-)\tau}{2}}. \tag{1}$$

Here we have phenomenologically included the characteristic decoherence time $T_d$, which includes the effects of oscillation relaxation and pure dephasing [16]. By fitting the data in Fig. 3c with Eq. (1), it is found that $T_d = 37.9$ ms. We also remark that the observed phenomena of the Rabi-like oscillations in few tens of mille-second indicate the coherence of mechanical oscillations is well preserved after the non-adiabatic transition.

**Classical analog of Stückelberg interferometry.** The investigations on the LZ transition and Rabi oscillations of the two-mode optomechanical system allow us to study the Stückelberg interferometry based on two subsequent passages through the avoided crossing. Specifically, as illustrated in Fig. 4a, after exciting the in-phase mode at $P_0 = 55.0\ \mu\text{W}$, we linearly ramp down the laser power to $P_f = 0.15\ \mu\text{W}$ in time $t_f$ such that the avoided crossing is passed through. This process acts as a tunable "beam splitter" which, depending on a sweeping rate $v$, allows adiabatic transfer, coherent splitting and diabatic transfer for the oscillation state. In particular, as shown previously, a 50/50-splitting is achieved with $v/2\pi = 342$ kHz/s. After the ramp, the laser power is kept constant at $P_f$ for a variable hold time $\tau$. A differential phase $\phi_{\text{adia}}(\tau)$ between the lower and upper states (normal modes) is accumulated. Clearly, $\phi_{\text{adia}}$ linearly increases with the hold time $\tau$. The laser power is then ramped back up to $P_0$, and the second passage creates a new superposition state of the upper and lower states depending on $\phi_{\text{adia}}$. Here, in analog to a Mach-Zehnder interferometer, this final step coherently recombines the oscillations leading to the constructive or destructive interference of the oscillation state in the two cantilevers and thus to fringes as a function of $\tau$. Figure 4b shows the rescaled oscillation amplitudes $A_2(2t_f + \tau)/A_2(0)$ as a function of hold time $\tau$ for various sweeping rates. As can be seen, for the 50/50-splitting ratio we indeed obtain the complete constructive and destructive interference fringes.

Quantitatively, the Stückelberg interferometry can be understood in terms of the adiabatic-impulse model [25], which assumes that the system evolves adiabatically unless at the avoided crossing where the LZ transition occurs. Based on this theory, the rescaled oscillation amplitude of the cantilever 2 after the double-passage LZ transition takes the form

$$\frac{A_2(2t_f+\tau)}{A_2(0)} = e^{-(2t_f+\tau)/2T_d}\sqrt{1 - 4p_{LZ}(1 - p_{LZ})\cos^2\Phi_{St}}, \quad (2)$$

where $\Phi_{St} = \phi_{adia}(\tau) + \phi_{dia}$ is the Stückelberg phase with $\phi_{adia}(\tau) = \frac{1}{2}\int_{t_1}^{t_2}(\omega_+ - \omega_-)dt$ being the adiabatic phase accumulated during the adiabatic evolution between $t = t_1$ and $t = t_2$ and $\phi_{dia}$ the phase acquired during the non-adiabatic transitions. Here we have still phenomenologically included the decoherence effect. Quantum mechanically, it is shown that the non-adiabatic phase is $\phi_{dia} = \phi_S - \frac{\pi}{2}$ with $\phi_S = \delta(\ln\delta - 1) + \arg[\Gamma(1 - i\delta)] + \frac{\pi}{4}$ being the Stokes phase, here $\delta = \frac{\Delta^2}{4\upsilon}$ and $\Gamma(z)$ is the gamma function [34]. Mathematically, the Stokes phase originates from the matching of the asymptotic forms of the Weber's function at a regular singularity [35]. Since $\phi_{adia}(\tau)$ can be evaluated analytically for the given experimental parameters, we may, as shown in Fig. 4b, determine $\phi_{dia}$ by fitting the measured data with Eq. (2). In Fig. 4c, we present the experimentally obtained non-adiabatic phases for various sweeping rates, which are in good agreement with those calculated using the Stokes phase.

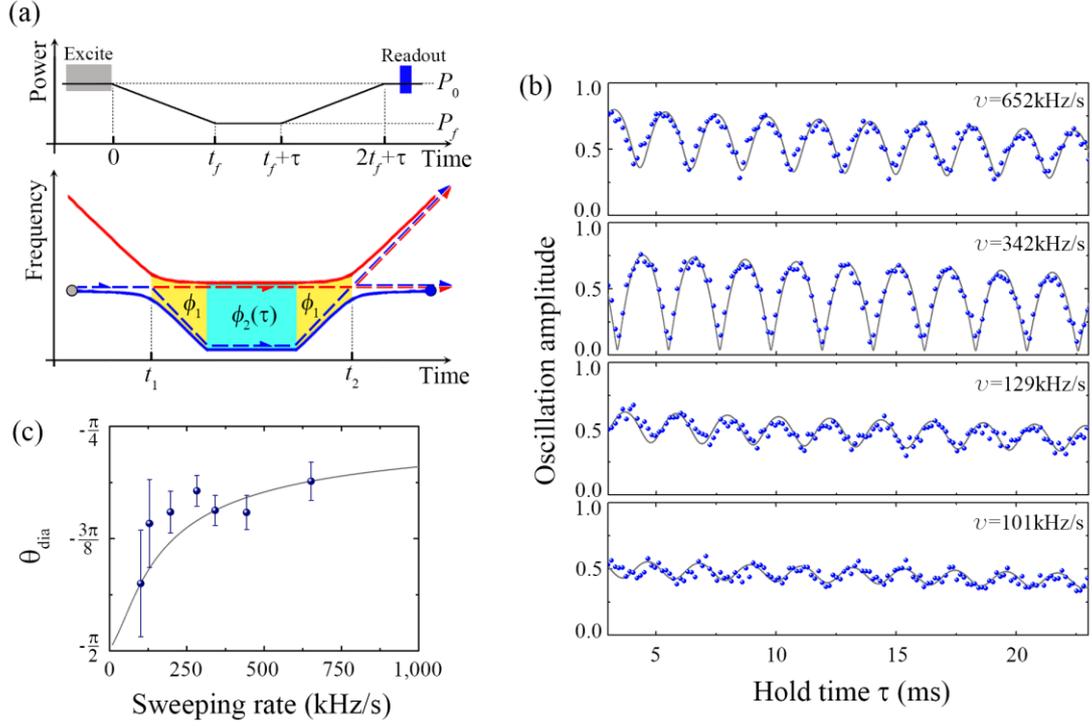

**Figure 4 | Stückelberg interferometry of mechanical oscillations.** (a) Optical driving scheme and the corresponding adiabatic energy levels. After the initial excitation for the in-phase mode far away from the avoided crossing, the system is driven through the avoided crossing twice. The final mechanical oscillations are then determined as a function of the acquired adiabatic phase $\phi_{\text{adia}}(\tau) = \phi_1 + \phi_2(\tau) + \phi_1$ for the upper and lower states (normal modes). (b) Interferometer fringes for various sweeping rates. The oscillation amplitude of the cantilever 2 is rescaled to its amplitude immediately after the initialization. Measured data are shown as dots and the fit with Eq. (2) is plotted as solid lines. (c) Non-adiabatic phases for various sweeping rates. Experimental results (dots) obtained from fitting of the Stückelberg oscillation in (b) are plotted with calculated results (solid line) using the formula of Stokes phase in the text.

**Discussion**

Based on the studies of the LZ transition and Rabi oscillation of two coupled cantilevers, we have realized a classical analog of the Stückelberg interferometer and measured the Stokes phase. Unlike the adiabatic phase that is of the dynamic nature, the Stokes phase of the LZ transition represents the phase acquired during the non-adiabatic transition. With a single passage through the avoided crossing of a two-level system, i.e., the LZ transition, one can determine the occupation probability on each level; while the relative phase between them is often difficult to measure. Although the Rabi oscillation manifests the phase coherence in the dynamics of the coupled cantilevers, no non-adiabatic phase is involved. Therefore, the measurement of the Stokes phase unveils an in-depth analog between the two-mode system and quantum two-level one. Our study may pave the way to new interferometric techniques using classical devices.

## Methods

**Experimental details.** All experiments are performed in high vacuum of $2 \times 10^{-9}$ Torr at room temperature. For the initial state preparation, in order to overcome the cantilevers' thermal motions which have amplitudes around 1 nm at 298 K, we resonantly excite the system with the target mode for 0.45 s until a stable oscillation with amplitude around 90 nm is generated. Immediately after the initial state preparation, an arbitrary waveform generator (AWG) is programmed to drive the electro-optical modulator (EOM) such that the power of the pumping laser is linearly swept. The fast decay of the low-finesse cavity allows producing an optical pulse inside the cavity in few-nano-second length such that instantaneous change of optomechanical interaction can be achieved as the laser power is swept. As to the experimental measurements, the same laser beam applied for optical pumping is also used to measure the oscillations of the coupled two cantilevers. Specifically, the motion of the cantilever 1 is monitored by measuring the intensity of the reflected light from the cavity with a photo detector (PD) after a 10 db beam splitter (BS). The magnitudes of the signal at the frequencies $\omega_+$ and $\omega_-$ are subsequently measured by lock-in amplifiers (LIAs) for the oscillation amplitudes of both the cantilevers. For a better measurement sensitivity, we always choose the final state at the right-hand side of the avoided crossing which corresponds to a higher laser power. In addition, to avoid the transient artifacts in the measurements, the final states in all our experiments are recorded 15 ms after the optical control sequence.

**Acknowledgements**

This work was supported by the National Basic Research Program of China (Grant No. 2012CB922104), and National Natural Science Foundation of China (Grant No. 11121403, 11434011, 11422437, and 11204357). We thank Wen-xian Zhang, Li-bin Fu and Yan-jun Ji for helpful discussions and Jia-dong Li for fabricating the coupled cantilevers.

Note added: During the preparation of the manuscript, we become award of the work of Seitner *et al.* [arXiv:1602.01034v1], in which the classical Stückelberg interferometry was realized with a two-mode electromechanical system.

**Author contributions**

H. F. and G.-Y. C. designed and conceived the experiment. H. F., Z.-C. G. and T.-H. M. carried out the measurements. Y. L. and S. Y. provided the theoretical support and data analysis. H. F., Y. L. and S. Y. wrote the manuscript. C.-P. S. and G.-Y. C. supervised the project. All authors discussed the results and contributed equally to the understanding.

**Competing financial interests:** The authors declare no competing financial interests.

# Supplementary Information

I. Theoretical model

In the absence of the mechanical excitation, the equations of the motion of the two-mode optomechanical system under consideration can be expressed as

$$\begin{pmatrix} \frac{d^2}{dt^2} + \gamma_1 \frac{d}{dt} + \omega_1^2(P) & J \\ J & \frac{d^2}{dt^2} + \gamma_2 \frac{d}{dt} + \omega_2^2 \end{pmatrix} \begin{pmatrix} x_1 \\ x_2 \end{pmatrix} = \frac{1}{m} \begin{pmatrix} F_1 \\ F_2 \end{pmatrix},$$

(S1)

where $x_i$ is the displacement of the $i$-th ($i=1,2$) cantilever, $m$ is the mass for both the cantilevers, $\omega_2$ [$\omega_1(P)$] is the intrinsic (optically-trapped effective) frequency for the second (first) cantilever, $J$ is the coupling strength between the cantilevers, and $F_i$ is the thermal Brownian force.

Diagonalizing the above equations with neglecting the effects of the damping and noises, we can obtain the out-of-phase ($X_+$) and an in-phase ($X_-$) motions of the normal modes $\begin{pmatrix} X_+ \\ X_- \end{pmatrix} = U \begin{pmatrix} x_1 \\ x_2 \end{pmatrix}$, where the transformation matrix $U \equiv U(P)$ represents the contributions of the cantilevers to the normal modes with the explicit form

$$U = \begin{pmatrix} u_{+1} & u_{+2} \\ u_{-1} & u_{-2} \end{pmatrix} = \begin{pmatrix} \cos\frac{\theta}{2} & \sin\frac{\theta}{2} \\ -\sin\frac{\theta}{2} & \cos\frac{\theta}{2} \end{pmatrix},$$

(S2)

where $\theta$ satisfies $\tan\theta = \frac{2J}{\omega_1^2 - \omega_2^2}$.

With the above process, Eq. (S1) can be written in the normal-mode motions as

$$\begin{aligned}\ddot{X}_+ + \omega_+^2 X_+ &= 0 \\ \ddot{X}_- + \omega_-^2 X_- &= 0\end{aligned} \tag{S3}$$

where $\omega_+(P)$ and $\omega_-(P)$ are the eigen-frequencies of the normal modes and

$$\omega_\pm^2 = \frac{1}{2}\left(\omega_1^2 + \omega_2^2 \pm \sqrt{(\omega_1^2 - \omega_2^2)^2 + 4J^2}\right). \tag{S4}$$

Note that the minimal frequency splitting between the normal modes, $\Delta \coloneqq \varepsilon(P_{AC}) = \omega_+(P_{AC}) - \omega_-(P_{AC})$, happening at the energy avoided crossing with the laser power $P_{AC}$ and $\omega_1(P_{AC}) = \omega_2$, is approximately proportional to the coupling strength: $\Delta \approx J/\omega_2$ when $J \ll \omega_1^2 \approx \omega_2^2$. Thus, one has $\tan\theta \approx \frac{\Delta}{\omega_1(P) - \omega_2} = \frac{\Delta}{\varepsilon(P)}$.

For a fixed trapping laser power $P$ (which means $\omega_1$ and $\omega_\pm$ are time-independent), the time dependence of the normal modes can be generally expressed as

$$X_\pm(t) \equiv \frac{1}{2}\tilde{X}_\pm(t) + c.c. = \frac{1}{2}\tilde{A}_\pm e^{-i\omega_\pm t} + c.c., \tag{S5}$$

where $\tilde{X}_\pm(t) = \tilde{A}_\pm e^{-i\omega_\pm t}$ represent the corresponding positive-frequency components with $|\tilde{X}_\pm(t)| = |\tilde{A}_\pm| =: A_\pm$ being the (constant) amplitudes of motions of the normal modes. Namely, the corresponding dynamics of (the positive-frequency components of) the normal modes starting from the initial time instant $t_0 = 0$ is subject to a free evolution

$$\begin{pmatrix} \tilde{X}_+(t) \\ \tilde{X}_-(t) \end{pmatrix} = N(t) \begin{pmatrix} \tilde{X}_+(0) \\ \tilde{X}_-(0) \end{pmatrix} \tag{S6}$$

with the free-evolution matrix

$$N(t) = \begin{pmatrix} e^{-i\omega_+ t} & 0 \\ 0 & e^{-i\omega_- t} \end{pmatrix}. \tag{S7}$$

Similarly, the motions of the cantilevers can be given in the form

$$x_j(t) := \frac{1}{2}\tilde{x}_j(t) + c.c. := \frac{1}{2}\tilde{A}_j(t)e^{-i\omega_j t} + c.c., \tag{S8}$$

where $\tilde{x}_j(t) = \tilde{A}_j(t)e^{-i\omega_j t}$ (j=1,2) represents the corresponding positive-frequency components for the j-th cantilevers with $|\tilde{x}_j(t)| = |\tilde{A}_j(t)| =: A_j(t)$ being the (time-dependent) amplitudes of motions of the j-th cantilever (when the coupling strength is much weak compared with the eigenfrequencies, $J \ll \omega_{1,2}^2$, which is fulfilled for most cases of coupled resonators). Accordingly, the solutions for the motions of the cantilevers can be given as

$$x_j(t) = \frac{1}{2}(u_{j+}\tilde{A}_+ e^{-i\omega_+ t} + u_{j-}\tilde{A}_- e^{-i\omega_- t}) + c.c. \tag{S9}$$

according to the solutions of the normal modes' motions.

We would like to point out that when the laser power $P$ is time-dependent, the above (constant) amplitudes for the normal modes, $A_\pm$, would be time dependent as $A_\pm(t)$. Meanwhile the above $N(t)$ would be generalized to be the form

$$N(t, t_0) = \begin{pmatrix} e^{-i\int_{t_0}^{t} \omega_+(t')dt'} & 0 \\ 0 & e^{-i\int_{t_0}^{t} \omega_-(t')dt'} \end{pmatrix}. \tag{S10}$$

## II. Rabi-like oscillations

As demonstrated by Fig. 3 and its related description, when the system is held in the coupling region with the fixed $P_f$ after the initial preparation, the coupling between the cantilevers allow coherent mechanical transferring from the cantilevers 2 to 1 following the Rabi-like oscillations. Such a oscillation starting from $t=0$ can be described analytically by the evolutions of the motions of $x_{1,2}(t) \equiv \frac{1}{2}\tilde{x}_{1,2}(t) + c.c.$ with the positive-frequency components satisfying

$$\begin{pmatrix}\tilde{x}_1(t)\\ \tilde{x}_2(t)\end{pmatrix} = R(t)\begin{pmatrix}\tilde{x}_1(0)\\ \tilde{x}_2(0)\end{pmatrix}. \tag{S11}$$

Here the newly defined operators

$$R(t) = U^{-1}(P_t)N(t)U(P_t)$$

$$= \begin{pmatrix} \cos^2\frac{\theta}{2}e^{-i\omega_+ t} + \sin^2\frac{\theta}{2}e^{-i\omega_- t} & \cos\frac{\theta}{2}\sin\frac{\theta}{2}(e^{-i\omega_- t} - e^{-i\omega_+ t}) \\ \cos\frac{\theta}{2}\sin\frac{\theta}{2}(e^{-i\omega_- t} - e^{-i\omega_+ t}) & \sin^2\frac{\theta}{2}e^{-i\omega_+ t} + \cos^2\frac{\theta}{2}e^{-i\omega_- t} \end{pmatrix}$$

(S12)

represents the operation of the Rabi-like oscillations for the positive-frequency parts of the motions of the cantilevers with $U(P_t)$ the corresponding mode function. And $N(t) = \begin{pmatrix} e^{-i\omega_+ t} & 0 \\ 0 & e^{i\omega_- t} \end{pmatrix}$ denotes the free evolution for the normal-mode basis $\{X_\pm\}$ for a fixed laser power. Namely, the evolution of the mechanical oscillations transferring between the cantilevers in the coupling region can be resorted to the free evolution for the normal modes sandwiched by the two related mode-function transformations.

In the case of Fig. 3, after the excitation for the cantilever 2 (in-phase normal mode), the trapping power is ramped linearly to $P_f$ in a very short time interval with the fast sweeping limit. That means the initial values for the Rabi-like oscillation of the cantilevers are approximately the excitation ones: $|\tilde{x}_2(0)| = A_2(0) \gg |\tilde{x}_1(0)| = A_1(0) \approx 0$. Thus, the (normalized) oscillation amplitudes of the cantilevers after the hold time $\tau$, are described analytically as

$$\frac{A_1(\tau)}{A_2(0)} = \frac{|\tilde{x}_1(\tau)|}{|\tilde{x}_2(0)|} = \left|\sin\theta \sin\frac{(\omega_+ - \omega_-)\tau}{2}\right|,$$

$$\frac{A_2(\tau)}{A_2(0)} = \frac{|\tilde{x}_2(\tau)|}{|\tilde{x}_2(0)|} = \sqrt{1 - \sin^2\theta \sin^2\frac{(\omega_+ - \omega_-)\tau}{2}}. \tag{S13}$$

That means the related Rabi frequency for the Rabi oscillations is $\Omega_R = \omega_+ - \omega_- = \sqrt{\Delta^2 + \varepsilon^2}$.

## III. Stückelberg interferometry

According to the adiabatic-impulse model [S1], the Stückelberg interferometry includes three processes: the first Landau-Zener transition, the adiabatical evolution, and the second Landau-Zener transition. In each transition mimicing a coherent oscillation-splitter, the transition matrix is given in the positive-frequency part of the normal-mode basis $\{X_\pm\}$ as

$$T = \begin{pmatrix} \sqrt{1-p_{LZ}}\, e^{-i\phi_{dia}} & -\sqrt{p_{LZ}} \\ \sqrt{p_{LZ}} & \sqrt{1-p_{LZ}}\, e^{i\phi_{dia}} \end{pmatrix} \qquad (S14)$$

with $p_{LZ}$ the transition probability and $\phi_{dia}$ the phase acquired during the non-adiabatic transitions (with their explicit forms given in the main text). During the adiabatic process between the two Landau-Zener transitions, the motions of the normal modes is governed by

$$N(t_2, t_1) = \begin{pmatrix} e^{-i\phi_+(\tau)} & 0 \\ 0 & e^{-i\phi_-(\tau)} \end{pmatrix} \qquad (S15)$$

with $\phi_\pm := \int_{t_1}^{t_2} \omega_\pm(t)dt$.

Accordingly, one can readily obtain the following motions of the positive-frequency part of the normal modes

$$\begin{pmatrix} \tilde{X}_+(2t_f + \tau) \\ \tilde{X}_-(2t_f + \tau) \end{pmatrix} = S \begin{pmatrix} \tilde{X}_+(0) \\ \tilde{X}_-(0) \end{pmatrix} \qquad (S16)$$

with

$$S = TN(t_2, t_1)T := \begin{pmatrix} S_{11} & S_{12} \\ S_{21} & S_{22} \end{pmatrix} \qquad (S17)$$

$$= \begin{pmatrix} (1-p_{LZ})e^{-i(2\phi_{dia}+\phi_+)} - p_{LZ}e^{-i\phi_-} & -\sqrt{p_{LZ}(1-p_{LZ})}[e^{-i(\phi_{dia}+\phi_+)} + e^{i(\phi_{dia}-\phi_-)}] \\ \sqrt{p_{LZ}(1-p_{LZ})}[e^{-i(\phi_{dia}+\phi_+)} + e^{i(\phi_{dia}-\phi_-)}] & (1-p_{LZ})e^{i(2\phi_{dia}-\phi_-)} - p_{LZ}e^{-i\phi_+} \end{pmatrix}.$$

Using the conditions $\tilde{x}_1(t) \approx \tilde{X}_+(t), \tilde{x}_2(t) \approx \tilde{X}_-(t)$ for the initial time instant $t = 0$ and the final one $t = 2t_f + \tau$ (where the system is far away from the avoided crossing) and the initial amplitude conditions $A_2(0) = |\tilde{x}_2(0)| \gg A_1(0) = |\tilde{x}_1(0)| \approx 0$, one can get the (normalized) final amplitudes of the cantilevers as

$$\frac{A_1(2t_f+\tau)}{A_2(0)} = \frac{|\tilde{x}_1(2t_f+\tau)|}{|\tilde{x}_2(0)|} = |S_{12}| = 2\sqrt{p_{\text{LZ}}(1-p_{\text{LZ}})}|\cos\Phi_{\text{St}}|,$$

$$\frac{A_2(2t_f+\tau)}{A_2(0)} = \frac{|\tilde{x}_2(2t_f+\tau)|}{|\tilde{x}_2(0)|} = |S_{22}| = \sqrt{1-4p_{\text{LZ}}(1-p_{\text{LZ}})\cos^2\Phi_{\text{St}}}, \tag{S18}$$

where $\Phi_{\text{St}} = \phi_{\text{dia}} + \phi_{\text{adia}}(\tau)$ with $\phi_{\text{adia}}(\tau) = \frac{1}{2}[\phi_+(\tau) - \phi_-(\tau)] = \frac{1}{2}\int_{t_1}^{t_2}[\omega_+(t) - \omega_-(t)]dt$ being the adiabatic phase accumulated during the adiabatic evolution. As shown in Fig. 4a, the adiabatic phase $\phi_{\text{adia}}(\tau)$ can be described by the following form $\phi_{\text{adia}}(\tau) = \phi_1 + \phi_2(\tau) + \phi_1$, where $\phi_1 = \frac{1}{2}\int_{t_1}^{t_f}[\omega_+(t) - \omega_-(t)]dt$ is the hold-time-$\tau$-independent phase and $\phi_2 = \frac{1}{2}\int_{t_f}^{t_f+\tau}[\omega_+(t) - \omega_-(t)]dt = \frac{1}{2}[\omega_+(P_f) - \omega_-(P_f)]\tau$ is the relevant phase.

## IV. Numerical Simulations

Now we investigate the numerical simulations of the time evolutions for the system of two coupled resonators under consideration by including the effects of the damping and thermal noises. To this aim, we rewrite the equations of the motion (S1) in the form as

$$\dot{x}_1 = \frac{p_1}{m},$$

$$\dot{p}_1 = -m\omega_1^2(P)x_1 - mJx_2 - \gamma p_1 + F_1(t),$$

$$\dot{x}_2 = \frac{p_2}{m},$$

$$\dot{p}_2 = -m\omega_2^2 x_2 - mJx_1 - \gamma p_2 + F_2(t), \tag{S19}$$

where $x_{1,2}$, $p_{1,2}$, $m$, $\omega_{1,2}$ and $\gamma$ are respectively the positions, momenta, masses, eigenfrequecies, and damping rates of the mechanical resonators. $J$ is the coupling strength between the resonators. Here the thermal Brownian noises $F_{1,2}(t)$ satisfy [S2]

$$\langle F_j(t)F_{j'}(t')\rangle = 2k_B T m \gamma \delta_{jj'} \delta(t-t') \tag{S20}$$

($\langle ... \rangle$ denotes the ensemble average) in the high-temperature limit $k_B T/(\hbar\omega_{1,2}) \gg 1$ as in our experiment. Here $k_B$, $T$, and $\hbar$ are the Boltzmann constant, ambient temperature, and reduced Plank constant, respectively.

In order to include the effects of both the damping and the thermal Brownian noises, here we consider the evolution of the close equations of motion for the mean values of the quadratic terms like $x_1^2$. Using the equations of motion in Eq. (S19) and the correlation

$\langle x_j(t) F_{j'}(t') \rangle = 0$,

$$\langle p_j(t) F_{j'}(t) \rangle = \begin{cases} k_B T m \gamma \delta_{jj'}, & (t = t') \\ 0, & (t \neq t') \end{cases}, \tag{S21}$$

we can get the following equations of motion

$$\frac{\partial}{\partial t}\boldsymbol{\mu} = -\mathbf{M}\,\boldsymbol{\mu} + \boldsymbol{\nu} \tag{S22}$$

with $\boldsymbol{\mu} = [\langle x_1^2\rangle, \langle x_1 p_1\rangle, \langle p_1^2\rangle, \langle x_2^2\rangle, \langle x_2 p_2\rangle, \langle p_2^2\rangle, \langle x_1 p_2\rangle, \langle x_2 p_1\rangle, \langle x_1 x_2\rangle, \langle p_1 p_2\rangle]^{\mathrm{T}}$,

$$\mathbf{M} = \begin{pmatrix} 0 & \frac{-2}{m} & 0 & 0 & 0 & 0 & 0 & 0 & 0 & 0 \\ m\omega_1^2 & \gamma & \frac{-1}{m} & 0 & 0 & 0 & 0 & 0 & mJ & 0 \\ 0 & 2m\omega_1^2 & 2\gamma & 0 & 0 & 0 & 0 & 2mJ & 0 & 0 \\ 0 & 0 & 0 & 0 & \frac{-2}{m} & 0 & 0 & 0 & 0 & 0 \\ 0 & 0 & 0 & m\omega_2^2 & \gamma & \frac{-1}{m} & 0 & 0 & mJ & 0 \\ 0 & 0 & 0 & 0 & 2m\omega_2^2 & 2\gamma & 2mJ & 0 & 0 & \frac{-1}{m} \\ mJ & 0 & 0 & 0 & 0 & 0 & \gamma & 0 & m\omega_2^2 & \frac{-1}{m} \\ 0 & 0 & 0 & mJ & 0 & 0 & 0 & \gamma & m\omega_1^2 & \frac{-1}{m} \\ 0 & 0 & 0 & 0 & 0 & 0 & \frac{-1}{m} & \frac{-1}{m} & 0 & 0 \\ 0 & mJ & 0 & 0 & mJ & 0 & m\omega_1^2 & m\omega_2^2 & 0 & 2\gamma \end{pmatrix}$$

and $\mathbf{v} = (0, 0, 2m\gamma k_B T, 0, 0, 2m\gamma k_B T, 0, 0, 0, 0)^T$.

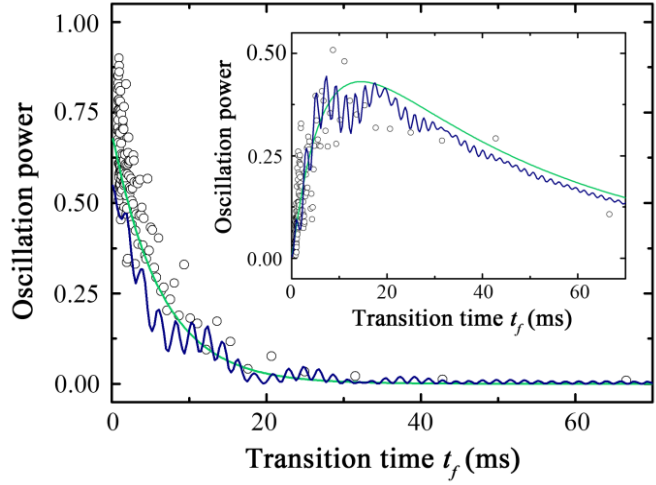

**Figure S1 | Oscillation power of cantilevers.** The oscillation powers of the cantilever 2 and (inset) cantilever 1 are rescaled to the oscillation power of the cantilever 2 immediately after the initialization. Experimental results (black circles) and the results calculated from the standard LZ formula considering the mechanical damping (green or light lines) are plotted in comparison with the related numerical results (blue or dark lines). Here the numerical calculations are performed with $m$=1.03 ng, $T$=298 K, and all the other parameters the same as those in Fig. 2.

According to the above equations (S22), we can obtain numerically the motion of each cantilever at any time instant once we know the initial conditions. For example, in the case of the Landau-Zener transition as given in Fig. 2, the corresponding initial condition is $\boldsymbol{\mu}(\mathbf{0}) = (A^2, 0, 2mk_BT, 2k_BT/(m\omega_2^2), 0, 2mk_BT, 0, 0, 0, 0)^{\mathrm{T}}$. The related numerical results of the oscillation energies (or the oscillation powers) are in good agreement with the experimental ones and the ones according to the formula of standard Landau-Zener transition, as seen in Fig. S1.

**Supplementary References**

S1. Shevchenko, S. N., Ashhab, S. & Nori, F. Landau–Zener–Stückelberg interferometry. *Phys. Rep.* **492**, 1-30 (2010).

S2. Gardiner, C. W. *Handbook of stochastic methods* 3rd edn (Springer, 2004).